\documentclass[letter,superscriptaddress,twocolumn,prl,showkeys,showpacs]{revtex4-1}

\usepackage[utf8]{inputenc}
\usepackage{textcomp}
\usepackage{amsmath}
\usepackage{graphicx}
\usepackage{verbatim}
\usepackage{color}
\usepackage{subfigure}
\usepackage{hyperref}
\usepackage{natbib}
\usepackage{ gensymb }
\usepackage{xfrac}

\begin{document}

\title{STM microscopy of the CDW in 1$T$-TiSe$_2$ in the presence of single atom defects}

\author{A. M. Novello}
\altaffiliation{Corresponding author.\\ anna.novello@unige.ch}
\affiliation{Department of Quantum Matter Physics, University of Geneva, 24 Quai Ernest-Ansermet, 1211 Geneva 4, Switzerland}

\author{B. Hildebrand}
\altaffiliation{Corresponding author.\\ baptiste.hildebrand@unifr.ch}
\affiliation{D{\'e}partement de Physique and Fribourg Center for Nanomaterials, Universit{\'e} de Fribourg, CH-1700 Fribourg, Switzerland}

\author{A. Scarfato}
\affiliation{Department of Quantum Matter Physics, University of Geneva, 24 Quai Ernest-Ansermet, 1211 Geneva 4, Switzerland}

\author{C. Didiot}
\affiliation{D{\'e}partement de Physique and Fribourg Center for Nanomaterials, Universit{\'e} de Fribourg, CH-1700 Fribourg, Switzerland}

\author{G. Monney}
\affiliation{D{\'e}partement de Physique and Fribourg Center for Nanomaterials, Universit{\'e} de Fribourg, CH-1700 Fribourg, Switzerland}

\author{A. Ubaldini}
\affiliation{Department of Quantum Matter Physics, University of Geneva, 24 Quai Ernest-Ansermet, 1211 Geneva 4, Switzerland}

\author{H. Berger}
\affiliation{Institut de G{\'e}nie Atomique, Ecole Polytechnique F{\'e}d{\'e}rale de Lausanne, CH-1015 Lausanne, Switzerland}

\author{D. R. Bowler}
\affiliation{London Centre for Nanotechnology and Department of Physics and Astronomy,
University College London, London WC1E 6BT, UK}

\author{P. Aebi}
\affiliation{D{\'e}partement de Physique and Fribourg Center for Nanomaterials, Universit{\'e} de Fribourg, CH-1700 Fribourg, Switzerland}

\author{Ch. Renner}
\affiliation{Department of Quantum Matter Physics, University of Geneva, 24 Quai Ernest-Ansermet, 1211 Geneva 4, Switzerland}

\begin{abstract}
We present a detailed low temperature scanning tunneling microscopy study of the commensurate charge density wave (CDW) in 1$T$-TiSe$_2$ in the presence of single atom defects. We find no significant modification of the CDW lattice in single crystals with native defects concentrations where some bulk probes already measure substantial reductions in the CDW phase transition signature. Systematic analysis of STM micrographs combined with density functional theory modelling of atomic defect patterns indicate that the observed CDW modulation lies in the Se surface layer. The defect patterns clearly show there are no 2$H$-polytype inclusions in the CDW phase, as previously found at room temperature [Titov A.N. \textit{et al}, Phys. Sol. State \textbf{53}, 1073 (2011). They further provide an alternative explanation for the chiral Friedel oscillations recently reported in this compound [J. Ishioka \textit{et al.}, Phys. Rev. B \textbf{84}, 245125, (2011)].    

\end{abstract}
\date{\today}
\pacs{68.37.Ef, 71.15.Mb, 74.70.Xa, 73.20.Hb}
\maketitle

The transition metal dichalcogenide (TMD) 1$T$-TiSe$_2$ has kept the scientific community wondering about a number of its striking physical properties for more than four decades \cite{Salvo1976, Wilson1977,  Wilson1978, Hughes1977,  Morosan2006a, Rasch2008, Kusmartseva2009a}. 1$T$-TiSe$_2$ is a layered compound consisting of a hexagonal layer of Ti sandwiched between two hexagonal layers of Se to form Se-Ti-Se sandwiches that stack via weak Van-der-Waals (VdW) forces to form a single crystal. The band structure of 1$T$-TiSe$_2$, as determined by angle-resolved photoemission spectroscopy, consists primarily of a Se 4p-valence band at the $\Gamma$ point and a Ti 3d-conduction band at the L point of the Brillouin zone. But it is still debated whether it is a semiconductor or a semimetal with evidences claimed for both alternatives \cite{Rasch2008,Greenaway1965a,Pillo2000,Monney2015}.

Below $T_{CDW}\approx 202K$, 1$T$-TiSe$_2$ undergoes a second order phase transition into a commensurate charge density wave (CDW). A comprehensive theory of this CDW formation is yet to be developed. Two main mechanisms are currently considered, driven either by a Jahn-Teller distortion \cite{Hughes1977, Rossnagel2002} or an excitonic ground state \cite{Wilson1977, Pillo2000, Cercellier2007a, Monney2009a}. The CDW phase has been found to melt upon copper intercalation \cite{Morosan2006a} or when applying pressure \cite{Kusmartseva2009a}. In both instances, superconductivity develops in a dome shaped region around some optimal doping or optimal pressure, with a maximum critical temperature of 4.1K and 1.8K, respectively. More recently, chiral properties have been reported for the CDW in pristine and copper intercalated 1$T$-TiSe$_2$ based on polarized optical reflectometry and scanning tunneling microscopy (STM) \cite{Ishioka2010,Ishioka2011, Iavarone2012}. 

Here we focus on the CDW instability in 1$T$-TiSe$_2$ in the presence of native atomic scale defects. Past studies performed using macroscopic probes including resistivity, magnetic susceptibility and optical reflectivity have found atomic intercalation and substitution to be detrimental to the CDW \cite{Salvo1976, Krasavin1998}. This compound is usually non-stoichiometric with a strong correlation between increasing crystal growth temperature and Ti self-doping leading to the collapse of the CDW phase transition signature in temperature dependent resistivity measurements \cite{Salvo1976}. STM offers new opportunities in allowing the simultaneous mapping of individual single atom defects and the CDW in real space, as well as measuring the local density of states (LDOS) around the Fermi level by tunneling spectroscopy. This technique has revealed a distorted CDW superlattice in doped 1$T$-TaS$_2$ \cite{Wu1988, Lieber1991}, a one-dimensional CDW in calcium intercalated graphite \cite{Rahnejat2011} and a finite CDW amplitude in the vicinity of intrinsic defects in 2$H$-NbSe$_2$ well above the bulk T$_{CDW}$ \cite{Arguello2014}. These examples highlight the possibility to gain insight into the CDW phase and its formation mechanism by means of STM in the presence of atomic defects and impurities. 

1$T$-TiSe$_2$ single crystals were grown by iodine vapour transport and cleaved in-situ below 10$^{-7}$mbar at room temperature. All measurements were performed on crystals grown at 650\celsius$~$\kern-1ex, except for the micrograph in fig. 3j acquired on a crystal grown at 575\celsius$~$to better observe atomic features unrelated to intercalated Ti. Constant current STM micrographs were recorded at 4.7K using an Omicron LT-STM and a SPECS JT-STM, with bias voltage V$_{\text{bias}}$ applied to the sample. Base pressure was in both cases better than 5$\times$10$^{-11}$mbar. DFT model calculations were performed using the plane wave pseudo-potential code VASP \cite{kresse1996, kresse1993}, version 5.3.3. Projector-augmented waves \cite{kresse1999} were used with the PBE \cite{perdew1996} exchange correlation functional and plane wave cut-offs of 211eV ($1T$-TiSe$_2$, I substitutional) and 400eV (O).  The cell size of our model was 28.035\AA$\times$28.035\AA. The $1T$-TiSe$_2$ surface was modeled with two layers and the bottom Se layer fixed. A Monkhorst-Pack mesh with $2\times2\times1$ $k$-points was used to sample the Brillouin zone of the cell.  The parameters gave energy difference convergence better than 0.01 eV. During structural relaxations, a tolerance of 0.03eV/\AA\ was applied. STM images were generated using the Tersoff-Hamann approach \cite{Tersoff1983} in which the current $I(V)$ measured in STM is proportional to the integrated LDOS of the surface using the bSKAN code \cite{Hofer2003}.

\begin{figure}
\includegraphics[scale=1]{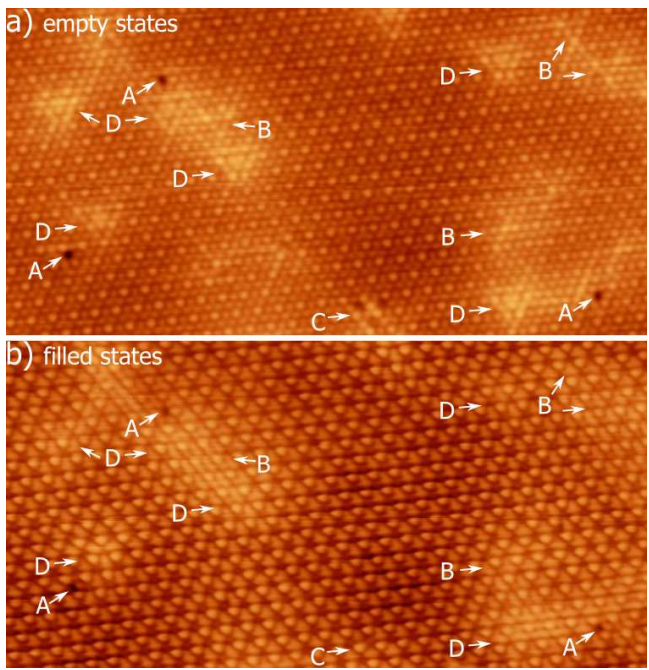}
\caption{\label{fig1}(color online). Simultaneously measured empty-states (a: V$_{\text{bias}}=$0.15V) and filled-states (b: V$_{\text{bias}}=-$0.15V) STM micrographs of a $1T$-TiSe$_2$ single crystal grown at 650\celsius. Image size: 22.2 nm $\times$11.4 nm, I$_{\text{t}}=0.2$nA, $T=4.7$K. Native defects are labelled A, B, C, D.}
\end{figure}

\begin{figure}
\includegraphics[scale=1]{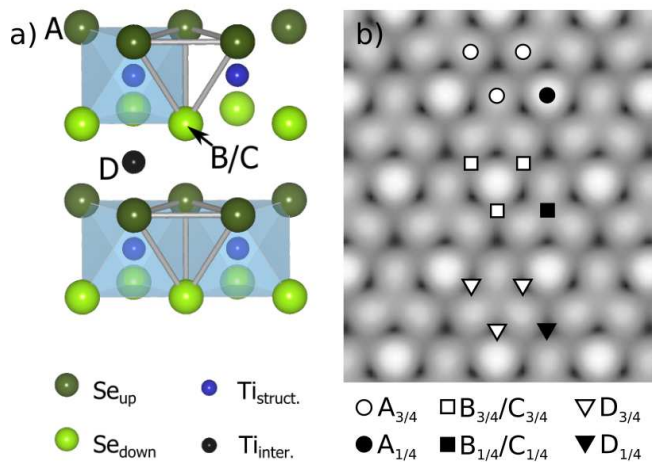}
\caption{\label{fig2}(color online). (a) Ball-and-stick model of the $1T$-TiSe$_2$ lattice showing the positions of the native defects A-D. (b) Model representation of the inequivalent $\sfrac{1}{4}$ (filled symbols) and $\sfrac{3}{4}$ (empty symbols) lattice sites for defects A-D in the commensurate CDW phase.}
\end{figure}

Fig. \ref{fig1} shows two high resolution STM micrographs of $1T$-TiSe$_2$ obtained at $T=4.7$K with exactly the same tip at positive and negative sample bias \footnote{Both image were taken simultaneously, positive bias being recorded while scanning the tip to the right and negative bias during the backward scan to the left.}. The bias voltages of $\pm150$mV have been chosen to enable the simultaneous resolution of the 2{a$_0$}$\times$2{b$_0$} CDW reconstruction on the selenium layer and atomic lattice features at opposite polarities. Defects (A-D) correspond to the dominant native atomic defects in $1T$-TiSe$_2$ identified in a recent STM/DFT study based on images recorded at a larger bias voltage where the CDW is not resolved \cite{Hildebrand2014}. These defects are Se surface vacancies (A), iodine (B) and oxygen (C) substitution for bulk Se and titanium intercalated into the VdW gap (D). Their positions in the lattice unit-cell are shown in Fig. \ref{fig2}(a). 

$1T$-TiSe$_2$ cleaves between the weakly VdW bonded Se-Ti-Se sandwiches, thus exposing a hexagonal Se layer to the surface. DFT modeling  enables us to identify the atomic lattice seen in STM maps with the Se surface layer by assigning the observed vacancies (defect A) to missing Se surface atoms \cite{Hildebrand2014}.
Thus, the commensurate in-plane 2{a$_0$}$\times$2{b$_0$} modulation [Fig. \ref{fig1}] is in perfect registry with the Se atomic lattice, indicating that the CDW charge modulation detected by STM resides in the Se layer.

As a consequence of the CDW modulation, there are two inequivalent sites in the unit-cell for each defect, with three times more $\sfrac{3}{4}$ than $\sfrac{1}{4}$ sites [Fig. \ref{fig2}(b)]. A survey of Se, O and Ti defects in a large area map (50$\times$50nm$^2$, 23'000 unit cells, 177 defects in total) yields approximately three times more $\sfrac{3}{4}$ than $\sfrac{1}{4}$ configurations for each of them. This uniform statistical distribution of all native defects among $\sfrac{3}{4}$ and $\sfrac{1}{4}$ sites implies they do not interact strongly with the CDW in this crystal, even though its resistive CDW transition is reduced by over 30\% compared to a sample with optimal stoichiometry. If they were interacting, we would expect dislocations to enable the CDW lattice to accommodate the random defect landscape. Indeed, we find no systematic domain formation, dislocations or weakening of the CDW lattice due to native defects. This rigidity of the CDW can be directly linked to its commensurate nature in $1T$-TiSe$_2$ \cite{McMillan1975}. In the same survey, we count about 80 intercalated Ti (defect D), corresponding to 0.35\% self doping, in excellent agreement with literature for samples grown at 650\celsius$~$\cite{Salvo1976}.  

\begin{figure*}
\includegraphics[scale=1]{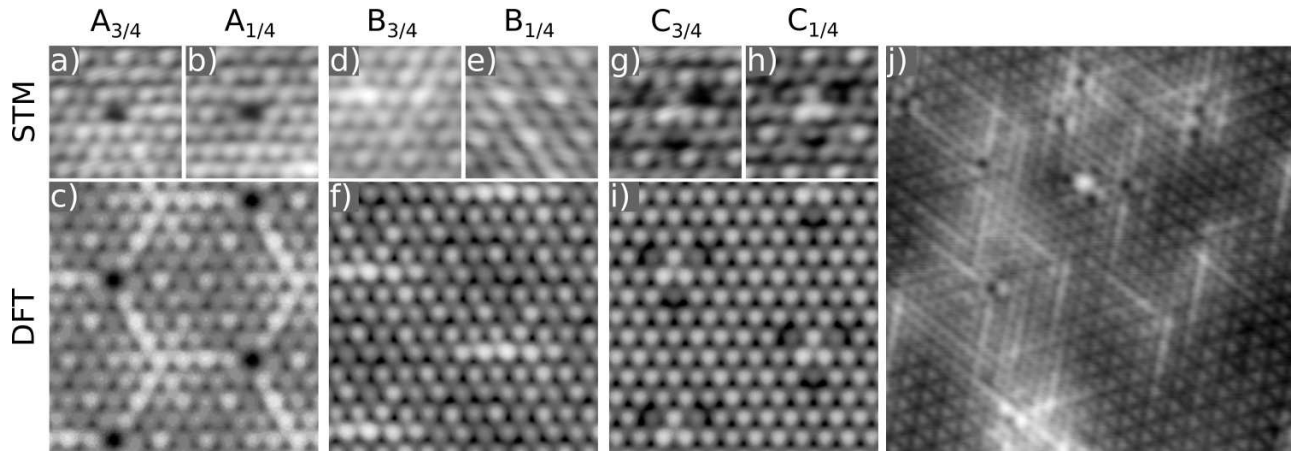}
\caption{\label{fig3}High resolution STM micrographs centered on Se vacancies (a,b), iodine (d,e) and oxygen (g,h) substitutions at $\sfrac{3}{4}$ sites (a,d,g) and $\sfrac{1}{4}$ sites (b,e,h) with corresponding DFT simulations calculated without including the CDW (c,f,i). (j) Linear features around defects A and B observed by STM on $1T$-TiSe$_2$ grown (11.5$\times$11.5nm$^2$, V$_{\text{bias}}=+150$mV, I$_{\text{t}}=0.2$nA). All images are taken on crystals grown at 650\celsius$~$except for image (j) recorded on a sample grown at 575\celsius$~$to have a reduced number of intercalated Ti.}
\end{figure*}

\begin{figure}
\includegraphics[scale=1]{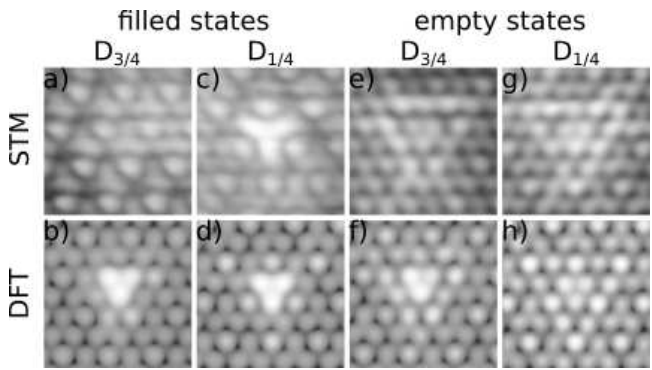}
\caption{\label{fig4} High resolution STM micrographs centered on intercalated Ti at $\sfrac{3}{4}$ sites (a,e) and $\sfrac{1}{4}$ sites (c,g) with corresponding DFT simulations in the presence of the CDW (b,d,f,h). V$_{\text{bias}}=-150$mV (a,b,c,d) and V$_{\text{bias}}=+150$mV (e,f,g,h), I$_{\text{t}}=0.2$nA.}
\end{figure}

Se vacancies appear as well resolved dark sites independent on bias voltage and position at the surface [Figs. \ref{fig3}(a)-(c)]. In contrast, defects B-D are mostly bright and best resolved and differentiated at positive V$_{\text{bias}}$ [Fig. \ref{fig1}]. Their characteristic patterns revealed by STM \cite{Hildebrand2014} are slightly modified in the presence of the CDW and depend on their $\sfrac{1}{4}$ or $\sfrac{3}{4}$ configuration (Figs. \ref{fig3} and \ref{fig4}). Of all defects, iodine substitution for Se$_{down}$ (defect B) is the most difficult to identify. On the $\sfrac{1}{4}$ site, it appears as a faint enhancement of the three nearest CDW maxima [Fig. \ref{fig3}(e)]. The $\sfrac{3}{4}$ configuration appears as a few atoms long brighter chain extending along one of the crystallographic direction [Fig. \ref{fig3}(d)]. Similar bright atomic chain features are found around defect A [Figs. \ref{fig3}(a) and (j)]. DFT simulations of defects A and B are in good agreement with these experimental observations [Figs. \ref{fig3}(c) and (f)]. The linear features are reproduced in the model without including the CDW instability. This shows they are not a different CDW ground state (e.g. 1D-CDW), but reflect local strain due to the Se vacancy and the larger atomic radius of iodine compared to Se. In regions with higher defect densities, these chains cooperate to form stripy patches in the STM topography, but without disrupting the long range coherent CDW [Fig. \ref{fig3}(j)]. The defects in these regions will produce an anisotropic deformation landscape explaining why these stripes do not always appear in all three high symmetry directions with the same intensity. When the defect density is low and in the vicinity of intercalated Ti (defect D), these stripes are usually less or not visible [Fig. \ref{fig1}(a)].

Defects C and D show more complex triangular patterns without the linear atomic features found around defects A and B. Oxygen substitution for Se$_{down}$ (defect C) is characterized by three bright central atoms centered on a larger, 60\degree$~$rotated triangle of three dark atoms [Figs.  \ref{fig3}(g) and (h)], in perfect agreement with DFT modeling  [Fig. \ref{fig3}(i)]. Titanium interstitials (defect D) appear as two concentric bright triangles centered on the defect [Fig. \ref{fig4}] \cite{Hildebrand2014}. The central triangles point in opposite directions in defect C compared to defect D. The triangular outline of defects C and D always point in the same direction in a given experiment [Fig. \ref{fig1}], attesting the perfect crystalline structure of our $1T$-TiSe$_2$ specimen. 
The unique triangle orientation and the perfect match between the data and the DFT models, which were all calculated in the 1$T$-polytype structure, imply there are no 2$H$-polytype inclusions where the coordination of the Ti atom changes from octahedral (1$T$) to trigonal-prismatic (2$H$) as found by Titov \textit{et al.} at room temperature \cite{Titov2011}. 
Although this finding cannot exclude a Jahn-Teller mechanism for the CDW origin \cite{Hughes1977,Rossnagel2002}, the capability to locally identify the 1$T$-phase may become instrumental in clarifying the role of local lattice modifications in the CDW formation.

The appearance of all native defects, except surface Se vacancies (defect A), change slightly depending on their $\sfrac{1}{4}$ or $\sfrac{3}{4}$ configuration. At positive sample bias, the oxygen substitution (defect C) in the $\sfrac{1}{4}$ configuration totally obscures the amplitude of the three nearest CDW maxima [Fig. \ref{fig3}(h)] whereas iodine on the same location (defect B) enhances them slightly [Fig. \ref{fig3}(e)]. Intercalated titanium (defect D) has an unmistakable triangular signature, with very sharp vertices in the $\sfrac{1}{4}$ configuration that become nearly extinct in the $\sfrac{3}{4}$ configuration. These different appearances of native atomic defects depending on their configuration ($\sfrac{1}{4}$ or $\sfrac{3}{4}$) suggest another explanation for the recently reported chiral Friedel oscillations. Our data and DFT modelling show the distinct left and right handed patterns discussed by Ishioka \textit{et al.} \cite{Ishioka2011} correspond in fact to different native defects (O and I-substitutions) in the two distinct $\sfrac{1}{4}$ and $\sfrac{3}{4}$ configurations, unrelated to chirality.

The native defects are poorly resolved in the negative low bias STM micrographs discussed here, except for intercalated Ti (defect D) in the $\sfrac{1}{4}$ configuration and Se vacancies (defect A). The dark sites associated with Se vacancies (defect A) correspond to \textit{holes} in the topography and are seen as such at both polarities. The other defect patterns are primarily electronic and their bias polarity dependent visibility observed here is consistent with a CDW gap that is biased towards occupied states at the Fermi level \cite{Iavarone2012}. A striking exception to this behavior is defect D which is nicely resolved in the $\sfrac{1}{4}$ configuration at V$_{\text{bias}}<0$ [Figs. \ref{fig1}(b) and \ref{fig4}(c)], closely matching the DFT modeling. The donor nature of intercalated Ti contributing electron states just above the occupied edge of the CDW gap \cite{Hildebrand2014} can explain the finite contrast of defect D at negative bias inside the CDW gap. However, it is presently not clear why only the $\sfrac{1}{4}$ configuration is resolved at V$_{\text{bias}}=-$150mV [Fig. \ref{fig4}(c)] while the $\sfrac{3}{4}$ configuration remains invisible [Fig. \ref{fig4}(a)]. This question demands further investigations, in particular in the context of a proposed excitonic ground state \cite{Wilson1977,Wilson1978}.

In summary, the careful analysis and comparison with DFT modeling  allows us to assign the surface patterns observed in STM micrographs of the CDW phase in $1T$-TiSe$_2$ exclusively to Se vacancies, O or I-substitutions and Ti intercalation \cite{Hildebrand2014}. We have shown the great potential of high resolution STM imaging of the CDW in the presence of atomic defects to gain insight into this ordered phase. We find that native defects have essentially no impact on the CDW lattice, at least up to the level of Ti self-doping considered here, where the corresponding phase transition is significantly reduced in transport measurements \cite{Salvo1976}. The only change, besides the characteristic signatures of each defect, is a locally enhanced brightness linked to their doping nature \cite{Hildebrand2014}. The theoretical proposal by McMillan \cite{McMillan1975} that atomic defects might trigger an incommensurate CDW is clearly not observed here.
For the first time, comparison with DFT modeling further allows us to unambiguously demonstrate that the observed defect patterns in the CDW phase are all consistent with the 1$T$-polytype, excluding 2$H$-polytype inclusions \cite{Titov2011}. Finally, our study sheds a different light on recently published work on the chiral nature of Friedel oscillations in the vicinity of defects \cite{Ishioka2011}. We find compelling evidence that the left and right-hand patterns identified in that work are the signatures of different $1T$-TiSe$_2$ native defects located on inequivalent lattice sites with respect to the CDW modulation.  

We thank H. Beck, F. Vanini and C. Monney for motivating discussions. Skillful technical assistance was provided by G. Manfrini, F. Bourqui, B. Hediger and O. Raetzo. This project was supported by the Fonds National Suisse pour la Recherche Scientifique through Div. II. A. M. Novello and B. Hildebrand contributed equally to this work.

\bibliography{library4}
\end{document}